\documentclass[aps,prd,a4paper,twocolumn,amsmath,showpacs,superscriptaddress,nofootinbib]{revtex4-1}

\usepackage{cancel}
\usepackage{cases}
\usepackage{subfigure}
\usepackage{graphicx}
\usepackage{dcolumn}
\usepackage{bm}
\usepackage{color}
\usepackage[colorlinks=true,pdfstartview=FitV,linkcolor=blue,citecolor=blue,urlcolor=blue]{hyperref}
\usepackage[mathlines]{lineno}
\usepackage[dvipsnames]{xcolor}
\usepackage{amsmath}
\usepackage{ytableau}
\everymath{\displaystyle}
\usepackage{textcase}
\usepackage{soul}

\enlargethispage{\baselineskip}

\everymath{\displaystyle}
\newcommand{\beq}{\begin{equation}}
\newcommand{\eeq}{\end{equation}}
\newcommand{\beqs}{\begin{eqnarray}}
\newcommand{\eeqs}{\end{eqnarray}}

\begin{document}
\title{Cosmic Birefringence and Electroweak Axion Dark Energy}

\author{Gongjun Choi}
\email{gongjun.choi@hotmail.com}
\affiliation{Tsung-Dao Lee Institute, Shanghai Jiao Tong University, Shanghai 200240, China}
\author{Weikang Lin}
\email{weikanglin@sjtu.edu.cn}
\affiliation{Tsung-Dao Lee Institute, Shanghai Jiao Tong University, Shanghai 200240, China}
\author{Luca Visinelli}
\email{luca.visinelli@sjtu.edu.cn}
\affiliation{Tsung-Dao Lee Institute, Shanghai Jiao Tong University, Shanghai 200240, China}
\affiliation{Istituto Nazionale di Fisica Nucleare, Laboratori Nazionali di Frascati, C.P. 13, I-100044 Frascati, Italy}
\author{Tsutomu T. Yanagida}
\email{tsutomu.tyanagida@sjtu.edu.cn}
\affiliation{Tsung-Dao Lee Institute, Shanghai Jiao Tong University, Shanghai 200240, China}
\affiliation{Kavli IPMU (WPI), UTIAS, The University of Tokyo, 5-1-5 Kashiwanoha, Kashiwa, Chiba 277-8583, Japan}
\date{\today}

\begin{abstract}
Taking the recently reported non-zero rotation angle of the cosmic microwave background (CMB) linear polarization $\beta=0.35\pm0.14{\rm\, deg}$ as the hint for a pseudo Nambu-Goldstone boson quintessence dark energy (DE), we study the electroweak (EW) axion quintessence DE model where the axion mass is generated by the EW instantons. We find that the observed value of $\beta$ implies a non-trivial $U(1)$ electromagnetic anomaly coefficient ($c_{\gamma}$), once the current constraint on the DE equation of state is also taken into account. With the aid of the hypothetical high energy structure of the model inspired by the experimentally inferred $c_{\gamma}$, the model is shown to be able to make prediction for the current equation of state ($w_{\rm DE,0}$) of the quintessence DE. This is expected to make our scenario distinguishable in comparison with the cosmological constant ($w=-1$) and testable in future when the error in the future measurement of $w_{\rm DE,0}$ is reduced to $\mathcal{O}(1)\%$ level ($\delta w=\mathcal{O}(10^{-2})$).
\end{abstract}

\maketitle

{\bf \textit{Introduction}\;---} The recently reported non-vanishing rotation angle $\beta=\mathcal{O}(0.1)\,$deg of the cosmic microwave background (CMB) linear polarization~\cite{Minami:2020odp} (a.k.a.\ cosmic birefringence) may serve as a crucial hint for the presence of new physics beyond the Standard Model (SM), pointing at a source for parity violation in Nature other than weak interactions. When the massive $\phi$ is coupled to the SM photon of tensor $F_{\mu\nu}$ via an operator $\propto (\phi/f_{\phi}) F_{\mu\nu}\tilde{F}^{\mu\nu}$, with $f_{\phi}$ a decay constant, a non-vanishing $\Delta\phi=\phi(t_{0})-\phi(t_{\rm LSS})$ can account for $\beta\neq0$~\cite{Minami:2020odp, Fujita:2020ecn, Takahashi:2020tqv, Fung:2021wbz, Nakagawa:2021nme, Jain:2021shf} with $\phi(t_{0})$ and $\phi(t_{\rm LSS})$ evaluated today and at the last scattering surface (LSS) respectively. For reasonable values of the coefficient of the electromagnetic anomaly, $\beta$ and $\Delta\phi/f_\phi$ have similar magnitude. 

Besides sourcing cosmic birefringence, the field $\phi$ might contribute to the energy density of the Universe, acting as a quintessence field~\cite{Ratra:1987rm, Zlatev:1998tr}. The simplest way to achieve this is to have $\phi$ evolve under a slow-roll motion to date, starting from an initial configuration close to $\pi f_\phi$ in field space. For $\phi$ to explain the dark energy (DE) observed, its mass must be as small as $m_{\phi}\simeq H_{0}\approx 10^{-33}{\rm\, eV}$. The natural candidate for this field is a Nambu-Goldstone boson (NGB) associated with a spontaneously broken global $U(1)_{X}$ symmetry~\cite{Fukugita:1994xn, Frieman:1995pm, Nomura:2000yk, Choi:1999xn}.
In this case, the equation of state (EoS) of $\phi$ ($w_{\phi}\equiv P_{\phi}/\rho_{\phi}$) is expected to lie in the range $-1\leq w_{\phi}\lesssim 0$, where $P_{\phi}$ and $\rho_{\phi}$ are the pressure and energy density of $\phi$ respectively. Therefore, the field $\phi$ can be identified with a candidate of the quintessence dark energy (QDE) characterized by the EoS today $w_{\rm DE,0}$.

Is it possible to realize a well-motivated and consistent QDE model in the particle physics which can account for $\beta=\mathcal{O}(0.1)\,$deg? For a non-Abelian $SU(N)$ gauge theory with the gauge coupling constant $g=\sqrt{4\pi\alpha}$, the instanton action ($S_{\rm inst}$) is given by $S_{\rm inst}=-2\pi n/\alpha(\rho)$ where $n$ and $\rho$ are the winding number and the size of the instanton. From the observation that $e^{-2\pi/\alpha(M_{P})}M_{P}^{4}\simeq\Lambda_{\rm DE}^{4}\simeq(2{\rm\, meV})^{4}$ where $M_{P}\simeq 2.4\times10^{18}{\rm\, GeV}$ is the reduced Planck mass, it was realized in Refs.~\cite{Fukugita:1994xn,Nomura:2000yk} that the electroweak (EW) $SU(2)$ instantons with $\rho\sim M_{P}^{-1}$ in the minimal supersymmetric extension of the SM (MSSM) can generate the axion potential comparable to the present energy density of DE. This led to the first proposal of the EW axion QDE model~\cite{Nomura:2000yk}. As will be shown later, however, it turns out that this original model cannot result in a sufficiently large $\beta$.

In this {\it Letter}, inspired by the integral observation of a non-zero $\beta$ and the charm of the model for the original EW axion QDE, we propose the extended, but more complete EW axion QDE model that is able to explain $\beta=\mathcal{O}(0.1)\,$deg. We derive the constraints on the model from the observation of the cosmic birefringence $\beta$ in Ref.~\cite{Minami:2020odp} and on the bound $w_{\rm DE,0}\lesssim-0.95$~\cite{Aghanim:2018eyx}. The model is based on MSSM with discrete $Z_{4R}$ $R$-symmetry, global $U(1)_{F}$ Froggatt-Nielsen symmetry and a global $U(1)_{X}$ symmetry anomalous with respect to the EW $SU(2)_{L}$. We shall examine how the two quantities ($w_{\rm DE}$ and $\beta$) are related to each other. In accordance with the theoretical criteria including the anomaly free $Z_{4R}$ and the perturbativity of $SU(2)_{L}\times U(1)_{Y}$ below the Planck scale, and the observational constraint on $\beta$ and $w_{\rm DE,0}$, we shall produce the model's prediction for $w_{\rm DE,0}$ and address how the next-generation cosmological observations test this EW axion QDE model against the case of a cosmological constant.

\begin{table*}[th]
\centering
\begin{tabular}{|c||c|c|c|c|c|c|c|c|c|c|c|c|c|c|c|c|c|c|c|c|} \hline
 & $Q$ & $\overline{U}$ & $\overline{D}$ & $L$ & $\overline{E}$ & $H_{u}$ &  $H_{d}$ &  $e^{-8\pi^{2}\tau}$ &  $\mathcal{D}^{2}$ &  $\overline{\mathcal{D}}^{2}$ & $\Psi$ & $\overline{\Psi}$ & $\Phi$ & $\overline{\Phi}$ & $X$ & $H_{u}'$ & $H_{d}'$ & $\Sigma$ & $\Sigma'$& $\epsilon\equiv\langle\phi\rangle/M_{P}$\\
\hline
$U(1)_{Y}$ & 1/6& -2/3& 1/3& -1/2& 1& 1/2& -1/2& -& -& -& -& -& -& -& -& 1/2& -1/2& -& -&-\\
\hline
$SU(2)_{L}$ & $\ytableausetup{textmode, centertableaux, boxsize=0.6em}
\begin{ytableau}
 \\
\end{ytableau}$ & - & - & $\ytableausetup{textmode, centertableaux, boxsize=0.6em}
\begin{ytableau}
 \\
\end{ytableau}$ & - & $\ytableausetup{textmode, centertableaux, boxsize=0.6em}
\begin{ytableau}
 \\
\end{ytableau}$ & $\ytableausetup{textmode, centertableaux, boxsize=0.6em}
\begin{ytableau}
 \\
\end{ytableau}$ & - & - & - &$\ytableausetup{textmode, centertableaux, boxsize=0.6em}
\begin{ytableau}
 \\
\end{ytableau}$ &$\ytableausetup{textmode, centertableaux, boxsize=0.6em}
\begin{ytableau}
 \\
\end{ytableau}$&-&-&-& $\ytableausetup{textmode, centertableaux, boxsize=0.6em}
\begin{ytableau}
 \\
\end{ytableau}$& $\ytableausetup{textmode, centertableaux, boxsize=0.6em}
\begin{ytableau}
 \\
\end{ytableau}$& ${\rm Ad}$&${\rm Ad}$&-\\
\hline
$Z_{4R}$      &  $3/5$  &  $3/5$  & $1/5$ & $1/5$ & $3/5$ & $4/5$ & $6/5$ & $-2$ & $-2$ & $+2$ & $1$ & $1$  & $0$ & $0$ & $+2$&$-6/5$&$6/5$&$0$&$+2$&-\\
\hline
$U(1)_{F}$& $(2,1,0)$& $(2,1,0)$&$(1,0,0)$ &$(1,0,0)$ &$(2,1,0)$ &- &- &$+10$ &- &- &- &- &- &- &-&-&-&-&-&-1\\
\hline
$U(1)_{X}$&-&-&-&-&-&-&-&-&-&-&-1&-&+1&-1&-&-&-&-&-&-\\
\hline
\end{tabular}
\caption{Representations and charge assignments under $U(1)_{Y}$, $SU(2)_{L}$, $Z_{4R}$, $U(1)_{F}$, and $U(1)_{X}$. The symbol “-" refers to the singlet. Each number in the parenthesis in the fifth row corresponds to $U(1)_{F}$ quantum numbers for each generation. The value of $\epsilon$ (last column) is the spurion used for suppressing the $U(1)_{\rm B+L}$ violating operator $\mathcal{O}_{\cancel{B+L}}=(QQQL)/M_{P}$.}
\label{table:qn} 
\end{table*}

{\bf \textit{EW Axion DE Model}\;---} With the particle content of the model given in Table~\ref{table:qn}, the guiding logic for the extension is:
\begin{enumerate}
\item Require a new discrete $R$-symmetry $Z_{4R}$ to be anomaly free via the introduction of a pair $(H_{u}',H_{d}')$;
\item Ensure that the axion dynamics does not suffer from a fine-tuned initial condition problem (achieved by introducing the $SU(2)_{L}$ triplet ($\Sigma,\Sigma'$) to make $\alpha_{2}(M_{P})\simeq1$)~\cite{Ibe:2018ffn};
\item Restrict the number of new heavy hyper-charged fields ($\Omega$) by demanding $U(1)_{Y}$ to remain perturbative up to the Planck scale.
\end{enumerate}

We first specify the symmetry group in the model. The interactions in the model respect\footnote{The gauged $Z_{4R}$ is assumed to originate from a global $U(1)_{R}$ which restricts all the tree-level interactions of the theory.} 
\beq
G=G_{\rm SM}\otimes Z_{4R}\otimes U(1)_{F}\otimes U(1)_{X}\,,
\label{eq:sym}
\eeq
where $G_{\rm SM}$ is the SM gauge symmetry group, $Z_{4R}$ is the gauged discrete $R$-symmetry,\footnote{$Z_{4R}$ is an appealing choice of a discrete $R$-symmetry in that it can be anomaly free gauged symmetry as shown in ~\cite{Ibanez:1991hv}.} $U(1)_{F}$ is the global flavor symmetry known as Froggatt-Nielsen symmetry and $U(1)_{X}$ is the global symmetry anomalous for $SU(2)_{L}$. Since  $Z_{4R}$ is assumed to be gauged, the cancellation of the mixed anomalies of $Z_{4R}\times[SU(2)_{L}]^{2}$ and $Z_{4R}\times[SU(3)_{c}]^{2}$ must be guaranteed. Note that massive chiral superfields transforming under $SU(2)_{L}$ and $SU(3)_{c}$ do not contribute to the anomaly~\cite{Ibanez:1991hv}. When only the MSSM particle contents are considered, the mixed anomalies read
\beqs
&&Z_{4R}\times[SU(2)_{L}]^{2}\rightarrow4+\left(-\frac{2}{5}\times3-\frac{4}{5}\right)\times3=-2\cr\cr
&&Z_{4R}\times[SU(3)_{c}]^{2}\rightarrow6+\left(-\frac{2}{5}\times2-\frac{2}{5}-\frac{4}{5}\right)\times3=0\,.\nonumber\\
\label{eq:anomaly}
\eeqs
According to Eq.~(\ref{eq:anomaly}), it becomes necessary for the model to introduce additional fields which can make the mixed anomaly $Z_{4R}\times[SU(2)_{L}]^{2}$ vanish. To this end, we introduce ($H_{u}',H_{d}'$), see Table~\ref{table:qn}, and thereby the mixed anomaly of $Z_{4R}\times[SU(2)_{L}]^{2}$ becomes 0 mod 4~\cite{Kurosawa:2001iq}.

Next, we investigate how the axion potential is generated to produce the required energy density for DE. To make $U(1)_{X}$ anomalous for $SU(2)_{L}$ and to induce the spontaneous breaking of $U(1)_{X}$, we introduce the chiral superfields $(\Psi,\overline{\Psi})$ and ($\Phi,\overline{\Phi}$), with $\langle\Phi\rangle=(F_{A}/\sqrt{2}){\rm exp}[\mathcal{A}/F_{A}]$ and $\langle\overline{\Phi}\rangle=(F_{A}/\sqrt{2}){\rm exp}[-\mathcal{A}/F_{A}]$. Here, the decay constant $F_{A}$ is the energy scale at which the spontaneous breaking of $U(1)_{X}$ occurs and $\mathcal{A}$ is the chiral superfield for the NGB associated with $U(1)_{X}$ with its scalar component ($S+iA$) composed of saxion ($S$) and axion ($A$). At this moment, only $\Psi(-1)$, $\Phi(+1)$ and $\overline{\Phi}(-1)$ are assumed to be charged under $U(1)_{X}$ with the corresponding charges specified in the parenthesis.

Thanks to this set-up, the part of the superpotential relevant to $(\Psi,\overline{\Psi})$, $(\Phi,\overline{\Phi})$, and $SU(2)_{L}$ gauge sector for the energy scale higher than $F_{A}$ reads
\beq
W\supset\frac{\tau}{4}\mathcal{W}^{a\alpha}\mathcal{W}^{a}_{\alpha}+\Phi\Psi\overline{\Psi}+X(\Phi\overline{\Phi}-2F_{A})\,,
\label{eq:W1}
\eeq
where we did not specify the dimensionless coupling constants, $a$ ($\alpha$) is the group (spinor) index, and the spurion superfield is
\beq
\tau=\frac{1}{g^{2}_{2}}+i\frac{\Theta}{8\pi^{2}}-\frac{2m_{1/2}}{g_{2}^{2}}\theta^{2}\,,
\label{eq:tau}
\eeq
where $g_{2}$ is the $SU(2)_{L}$ gauge coupling, $\Theta$ is the $SU(2)_{L}$ vacuum angle, $m_{1/2}$ is the soft SUSY-breaking mass for $SU(2)_{L}$ gauginos and $\theta$ is a Grassmann variable. In Eq.~(\ref{eq:W1}), the chiral superfield $X$ ensures the acquisition of the vacuum expectation value (VEV) $\langle\Phi\rangle=\langle\overline{\Phi}\rangle=F_{A}/\sqrt{2}$ with assumption of the same soft masses for $\Phi$ and $\overline{\Phi}$.

After $U(1)_{X}$ gets spontaneously broken, the coupling of $\mathcal{A}$ to the $SU(2)_{L}$ field strength arises,
\beq
\mathcal{L}\supset\int d^2\theta\left(\frac{1}{32\pi^2}\frac{\mathcal{A}}{F_{A}}\mathcal{W}^{a\alpha}\mathcal{W}^{a}_{\alpha} + {\rm h.c.}\right)\,,
\label{eq:LaWW}
\eeq
as it can be easily seen via the field redefinition of $\Psi$ to absorb ${\rm exp}[\mathcal{A}/F_{A}]$.
 
With the axion coupling to $SU(2)_{L}$ instanton in Eq.~(\ref{eq:LaWW}), $SU(2)_{L}$ instantons generate the axion potential and provide the axion with a mass. However, since $U(1)_{B+L}$ is well-known to be anomalous with respect to $SU(2)_{L}$ in MSSM, the axion becomes massless unless $U(1)_{B+L}$ is explicitly broken down~\cite{Nomura:2000yk}.\footnote{For $U(1)_{B+L}$, $B$ and $L$ stand for baryon and lepton number respectively.} In order to introduce a mass term for the axion, the dimension five operator $\mathcal{O}_{\cancel{B+L}}=(QQQL)/M_{P}$ explicitly breaking $U(1)_{B+L}$ is introduced in the model. Setting $\epsilon\simeq1/17$ for correctly producing the quark and lepton mass matrices~\cite{Buchmuller:1992qc}, the operator $\mathcal{O}_{\cancel{B+L}}$ can be sufficiently suppressed by powers of $\epsilon$ to avoid the potential inconsistent rapid proton decay~\cite{Sakai:1981pk, Weinberg:1981wj}.

The EW axion potential receives contributions from the effective K{\"a}hler potential ($K_{\rm eff}$) and effective superpotential ($W_{\rm eff}$) induced by the $SU(2)_{L}$ instanton for the energy scale below $\rho^{-1}=M_{P}$ with $\rho$ denoting the instanton size of interest. Terms in $e^{-8\pi^{2}\tau}K_{\rm eff}$ and $e^{-8\pi^{2}\tau}W_{\rm eff}$ are determined by symmetries in the model and holomorphy of the superpotential~\cite{Choi:1996fs, Choi:1998ep}. Importantly, supercovariant derivatives can appear only in $K_{\rm eff}$. Dominating over $W_{\rm eff}$ in contribution to the EW axion potential, $K_{\rm eff}$ includes $\overline{\mathcal{D}}^{2}$ and a net negative power of $\rho$ to explain $R$-charge and the mass dimension of $K_{\rm eff}$ respectively. As a result, $K_{\rm eff}$ is dominated by instantons of the size $\rho\simeq M_{P}^{-1}$ and eventually produces the EW axion potential~\cite{Nomura:2000yk} 
\beqs
{\Lambda_{\mathcal{A}}}^4&\simeq&c\, e^{-\frac{2\pi}{\alpha_2(M_{P})}}\epsilon^{10}m_{\rm SUSY}^3 M_{P}\cr\cr
&\times&\left(\frac{m_{3/2}}{M_{P}}\right)^{2T(\ytableausetup{textmode, centertableaux, boxsize=0.6em}
\begin{ytableau}
 \\
\end{ytableau})}\left(\frac{m_{\Psi}}{M_{P}}\right)^{2T(\ytableausetup{textmode, centertableaux, boxsize=0.6em}
\begin{ytableau}
 \\
\end{ytableau})}\left(\frac{m_{\Sigma}}{M_{P}}\right)^{2T({\rm Ad})}\cr\cr
&\simeq&c\, e^{-\frac{2\pi}{\alpha_{2,{\rm MSSM}}(M_{P})}}\epsilon^{10}m_{\rm SUSY}^3 M_{P}\cr\cr
&\simeq& c\epsilon^{10}(1{\rm\, eV})^{4}\,,
\label{eq:EWaxionpotential}
\eeqs
where $c$ is a combination of coefficients of higher dimensional operators used for closing MSSM fermion zero modes emitted from the EW instanton and $m_{\rm SUSY}=\mathcal{O}(1){\rm\, TeV}$ is the soft SUSY-breaking mass.\footnote{For communicating the SUSY-breaking in the invisible sector to the visible sector, we are assuming the gravity mediation~\cite{Nilles:1983ge}. See Ref.~\cite{Choi:2019jck} for the case of the gauge mediation.} In Eq.~(\ref{eq:EWaxionpotential}), the term $\epsilon^{10}$ compensates the total $U(1)_{F}$ charge of higher dimensional operators closing MSSM fermion zero modes. On the other hand, $m_{3/2}$, $m_{\Psi}$, and $m_{\Sigma}$ are the gravitino mass for $(H_{u}',H_{d}')$ and the bare masses of ($\Psi,\overline{\Psi}$) and ($\Sigma,\Sigma'$) respectively. Note that factors including these masses are fermion zero modes contribution of non-MSSM fields charged under $SU(2)_{L}$ and the modification to $\alpha_{2,{\rm MSSM}}(M_{P})=1/23$ due to these fields cancel the factors at the one-loop level (referred to as “SUSY-miracle")~\cite{Ibe:2018ffn}.\footnote{Factors including masses of the non-MSSM fields in Eq.~(\ref{eq:EWaxionpotential}) can be understood to be closing each $2T(R)$ pair of fermion zero modes of ($H_{u}',H_{d}'$), ($\Psi,\overline{\Psi}$) and ($\Sigma,\Sigma'$) using the mass insertion with $\rho m_{3/2}$, $\rho m_{\Psi}$, and $\rho m_{\Sigma}$. Here $T(R)$ is the Dynkin index of the $SU(2)_{L}$ representation of $R$ in which each field transforms.} This fact that ${\Lambda_{\mathcal{A}}}^{4}$ remains almost insensitive to the structure of the model deviating from MSSM is reflected in the second equality. Eventually through Eq.~(\ref{eq:EWaxionpotential}) the model is confirmed to be able to produce axion potential that can account for the energy density of DE.

As the last part of the study of the model, we examine the renormalization group evolution (RGE) of gauge couplings of $SU(2)_{L}$ and $U(1)_{Y}$. Defining the displacement of the axion field from the hilltop ($\delta A\equiv \pi F_{A}-A$), the initial $\delta A$ at the time satisfying $H>\!\!>m_{A}$ is required to be extremely fine-tuned unless $F_{A}$ is close to $M_{P}$. On the other hand, the weak gravity conjecture (WGC) setting the upper bound on the instanton action $S_{\rm inst}=2\pi/\alpha_{2}<M_{P}/F_{A}$~\cite{ArkaniHamed:2006dz}, the larger $F_{A}$ demands the larger $\alpha_{2}(M_{P})$. Thus, to avoid such a fine-tuning and to remain consistent with WGC, we shall consider the scenario of $F_{A}=\mathcal{O}(0.1)M_{P}$ with $\alpha_{2}(M_{P})\simeq1$. 

As a matter of fact, the main reason why we introduced two $SU(2)_{L}$ triplets ($\Sigma,\Sigma'$) in Table~\ref{table:qn} precisely lies in our intention to accomplish this scenario. We find that for $m_{\Sigma}=\mathcal{O}(10^{7}){\rm\, GeV}$, $\alpha_{2}(M_{P})\simeq1$ could be accomplished.\footnote{Of course not only $(\Sigma,\Sigma')$ but other non-MSSM fields charged under $SU(2)_{L}$ shown in Table~\ref{table:qn} contribute to RGE of $\alpha_{2}$. Nevertheless, introducing additional $SU(2)_{L}$ triplets are essential to have $\alpha_{2}(M_{P})\simeq1$. Note that even if we introduce fields other than $(\Sigma,\Sigma')$ to achieve $\alpha_{2}(M_{P})\simeq1$, the result in Eq.~(\ref{eq:EWaxionpotential}) does not change.} Although through Eq.~(\ref{eq:EWaxionpotential}) we already confirmed that the model's prediction ${\Lambda_{\mathcal{A}}}^{4}\simeq(2{\rm\, meV})^{4}$ remains unaffected by the change in $\alpha_{2}(M_{P})$ triggered by introduction of heavy particles charged under $SU(2)_{L}$~\cite{Ibe:2018ffn}, the change in RGE of $\alpha_{2}$ is expected to cause that of $\alpha_{1}=g_{Y}^{2}/(4\pi)$ where $g_{Y}$ is the gauge coupling constant of $U(1)_{Y}$. The RGE of $\alpha_{1}$ is affected by $\alpha_{2}$ at the two loop level, so that an increase in $\alpha_{2}$ leads to a larger beta function for $\alpha_{1}$.

The particle contents in Table~\ref{table:qn} shows that $\alpha_{1}(M_{P})\lesssim1$, so that the model presented is perturbative. Thus as far as the EW sector is concerned, the model lacks any misunderstanding possibly caused by a non-perturbative effect below the Planck scale. However, the model is phenomenologically required to have more fields ($\Omega$) charged under both $U(1)_{Y}$ and $U(1)_{X}$ to produce significant mixed anomaly of $U(1)_{X}\times[U(1)_{\rm EM}]^{2}$. For the model not only to explain the cosmic birefringence but also to maintain the perturbativity, the number of $\Omega$ should be large enough but still needs to be upper-bounded. We discuss this upper-bound and how it affects the model's prediction for EoS of quintessence DE in the next section.

{\bf \textit{EW Axion DE Phenomenology}\;---} In the model, apart from the EW axion potential generated by EW instantons, the presence of $\Psi$ induces the following inevitable interaction between the EW axion ($A$) and the SM photon,
\beq
\mathcal{L}_{\rm eff}\supset-c_{\gamma}\frac{g_{\rm em}^{2}}{16\pi^{2}}\frac{A}{F_{A}}F^{\mu\nu}\tilde{F}_{\mu\nu}+\frac{V_{0}}{2}\left[1-\cos\left(\frac{A}{F_{A}}\right)\right]\,,
\label{eq:AFF}
\eeq
where $g_{\rm em}$ is the gauge coupling constant for the electromagnetic $U(1)_{\rm em}$ gauge symmetry and $c_{\gamma}=1$ when only $\Psi$ contributes to the anomaly of $U(1)_{X}\times[U(1)_{\rm em}]^{2}$. 

\begin{table}[th]
    \centering
    \begin{tabular}{c||c|c|c|c|c|c|c|c|c}
    \hline
        $F_{A}/M_{P}$ & 0.3 & 0.25 & 0.2 & 0.15 & 0.1 & 0.075 & 0.05 & 0.025 & $\rightarrow0$ \\
        \hline
        $\xi$ & 0.85 & 0.82 & 0.80 & 0.81 & 0.85 & 0.89 & 0.93 & 0.96 & 1  \\
    \hline
    \end{tabular}
    \caption{Values of the factor ($\xi$) in Eq.~(\ref{eq:wDE}) found via the numerical analysis to relate $w_{\rm DE,0}$ and $\beta/c_{\gamma}$.}
    \label{table:factor}
\end{table}

Thanks to Eq.~(\ref{eq:AFF}), the recently reported non-vanishing rotation angle ($\beta=0.35\pm0.14{\rm\, deg}$) of the linearly polarized CMB can be equated to \cite{Takahashi:2020tqv}
\beq
\beta=0.42{\rm\, deg}\times \frac{c_{\gamma}}{2\pi}\times\frac{A(t_{0})-A(t_{\rm LSS})}{F_{A}}\,,
\label{eq:beta}
\eeq
where $c_{\gamma}\equiv\sum_{i=\Psi,\Omega}Q_{X,i}Q_{{\rm em},i}^{2}$ and $Q_{X}$, $Q_{\rm em}$ are the charges of $U(1)_{X}$ and $U(1)_{\rm em}$ respectively.

Intuitively, for the slow-rolling to be currently maintained, we expect $A$ to initially lie near the hilltop of the potential, i.e.\ $A_{\rm hill} =\pi F_{A}$, and to have not yet evolved near the inflection point $A_{\rm infl} = (\pi/2) F_{A}$, implying $\Delta A/F_{A}\equiv [A(t_{0})-A(t_{\rm LSS})]/F_{A} \lesssim 1$. For the model to produce $\beta$ as large as $\sim0.35\,$deg, a value of $c_{\gamma} \gtrsim 10$ is required. For this, we introduce additional chiral superfields $\Omega$ charged under both $U(1)_{Y}$ and $U(1)_{X}$, with the hypercharge $+1$ and $U(1)_{X}$ charge $+1$.\footnote{Along with $\Omega$, we also introduce $\overline{\Omega}$ which has $U(1)_{Y}$ charge -1, but serves as the singlet of $U(1)_{X}$. This set-up introduces the Yukawa coupling $W\supset y\Phi\Omega\overline{\Omega}$ to the superpotential and imposes the supersymmetric mass $m_{\Omega}=yF_{A}/\sqrt{2}$ to $\Omega$ supermultiplet when $U(1)_{X}$ is spontaneously broken. Here $y$ is a $\mathcal{O}(1)$ dimensionless coupling constant.} The fields $\Omega$ are assumed to be $SU(3)_{c}\otimes SU(2)_{L}$ singlets. For perturbativity of the theory, we demand $\alpha_{1}(M_{P})\lesssim1$,\footnote{To have the RGE of $\alpha_{2}$ reflected in that of $\alpha_{1}$, we use two-loop level beta function of $\alpha_{1}$.} which constrains the number of $\Omega$ fields. This in turn produces a maximum value $c_{\gamma,{\rm max}}$ for the anomaly coefficient. Using this theoretical constraint $c_{\gamma}<c_{\gamma,{\rm max}}$ and the observational constraints for $\beta$ and $w_{\rm DE,0}$, we derive the model's prediction for the EoS of the EW axion DE, which is the main issue for the study of DE properties.

\begin{figure}[t]
\centering
\hspace*{-5mm}
\includegraphics[width=0.48\textwidth]{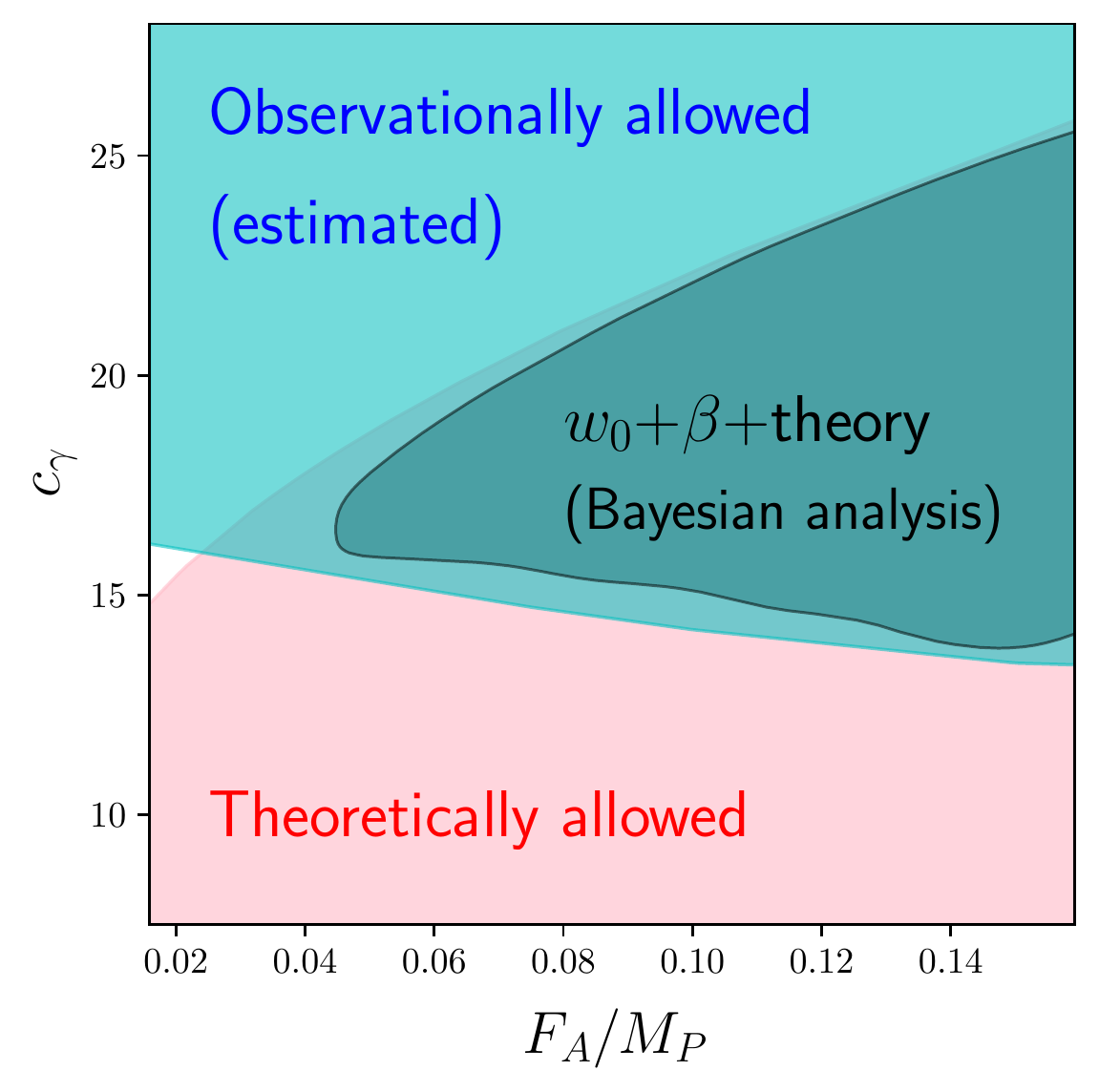}
\caption{The plot of the constraint on $(F_{A}/M_{P},c_{\gamma})$. For each $F_{A}/M_{P}$, the upper boundary shows $c_{\gamma,{\rm max}}$ obtained by requiring $\alpha_{1}(M_{P})\lesssim1$. The upper limit $F_A/M_P\lesssim 1/(2\pi)$ is set by the weak gravity conjecture. As the result of application of Eq.~(\ref{eq:wDE}), the blue region shows the estimated observationally allowed region satisfying $w_{\rm DE,0}<-0.95$ for $\beta=0.35$. The red region satisfies the theoretical bound $c_{\gamma}<c_{\gamma,{\rm max}}$. The black region shows the $1\sigma$ contour via a Bayesian analysis taking into account the theoretical bound $c_{\gamma}<c_{\gamma,{\rm max}}$ and $F_A$ as well as the observational constraints on $\beta$ and $w_{\rm DE,0}$.}
\vspace*{-1.5mm}
\label{fig:1}
\end{figure}

The time evolution of the EW axion obeys the following equation of motion
\beq
\ddot{A}+3H\dot{A}+\frac{\partial V(A)}{\partial A}=0\,,
\label{eq:eomofaxion}
\eeq
where $V(A)$ is the cosine-type function given in Eq.~(\ref{eq:AFF}), $H$ is the Hubble expansion rate, and a dot is a differentiation with respect to cosmic time. The evolution of the axion field depends on the set of parameters $(V_{0},F_{A})$ via the axion potential from Eq.~\eqref{eq:AFF}. Assuming a spatially flat universe and $\dot{A}=0$ initially, we numerically solve Eq.~(\ref{eq:eomofaxion}) in which $A$ also contributes to DE through $H$. By scanning over different sets of $(V_{0},F_{A})$, we obtain different solutions which in turn give us a list of today's EoS of the EW axion $w_{\rm DE,0}$ and $\beta/c_{\gamma}$, as well as the following relation between them,
\beq
w_{\rm DE,0}=-1+2\pi^2\xi^2\left(\frac{\beta/c_\gamma}{0.42\,\rm{deg}}\right)^2\,,
\label{eq:wDE}
\eeq
where $\xi$ is a numerical factor which is specified in Table~\ref{table:factor} as a function of $F_{A}/M_{P}$. Note that Eq.~(\ref{eq:wDE}) can be used for any DE model as far as the model has the same low energy effective Lagrangian as Eq.~(\ref{eq:AFF}).

Experimentally $\beta$ is constrained by the results in Ref.~\cite{Minami:2020odp}, while the likelihood analysis on the data combination {\it Planck} TT,TE,EE+lowE+lensing+BAO+SNe with a flat prior $w_{\rm DE,0}\geq -1$ leads to $w_{\rm DE,0}<-0.95$ at 95\% confidence level (C.L.)~\cite{Aghanim:2018eyx}. Applying these to Eq.~(\ref{eq:wDE}), we first estimate the experimentally allowed region on the plane $(F_{A}/M_{P},c_{\gamma})$. This is shown as the blue region in Fig.~\ref{fig:1}. We notice that $c_{\gamma} \gtrsim 10$ is required, which experimentally supports our intuitive understanding for the need to introduce additional $\Omega$ fields. On the other hand, setting $m_{\Omega}\equiv yF_{A}/\sqrt{2}$ as the mass of $\Omega$, we apply the perturbativity criterion ($\alpha_{1}(M_{P})\lesssim1$) to obtain $c_{\gamma,{\rm max}}$ as a function of $F_{A}/M_{P}$ and show the result in the red region of Fig.~\ref{fig:1}. Finally, we perform a Bayesian analysis taking into account both the theoretical bounds of $c_{\gamma}$, $F_A/M_P$, as well as observational constraints on $\beta$ and $w_{\rm DE,0}$. This leads to the joint constraint on $(F_{A}/M_{P},c_{\gamma})$ shown as the black region in Fig.~\ref{fig:1}.

The Bayesian analysis also yields the model's prediction for $w_{\rm DE,0}$, using Eq.~(\ref{eq:wDE}) and marginalizing over $\beta$, $c_{\gamma}$ and $F_A$. For the current precision of $\beta$ and upper bound of $w_{\rm ED,0}$, the model predicts $-0.994<w_{\rm DE,0}<-0.968$ (68\% C.L.).  With the near-future CMB missions, the statistical uncertainty of $\beta$ can be reduced by an order of magnitude \cite{Pogosian:2019jbt}. By then, an even tighter prediction $-0.982<w_{\rm DE,0}<-0.961$ (68\% C.L.) can be made by assuming the same mean value of $\beta$.\footnote{Even if the uncertainty of $\beta$ is only improved by a factor of four, the model already predicts $-0.983\!\!<w_{\rm DE,0}\!\!<-0.962$ (68\% C.L.).} This is exciting because the targeting error budget of $w_{\rm DE,0}$ is $\mathcal{O}(10^{-2})$ with the next-generation cosmological observations including Euclid CMB mission~\cite{Amendola:2016saw}, the Rubin observatory~\cite{Abate:2012za} and DESI~\cite{Aghamousa:2016zmz}. This means our DE scenario will be soon able to be distinguishable from the case of a cosmological constant and detected by observations.

{\bf \textit{Conclusions}\;---}We have presented a complete model for an EW axion quintessence dark energy, based on MSSM with discrete $Z_{4R}$ $R$-symmetry, $U(1)_{F}$ Froggatt-Nielsen symmetry and a global $U(1)_{X}$ symmetry anomalous with respect to the EW $SU(2)_{L}$. Projecting on the model the recent detection of the cosmic birefringence~\cite{Minami:2020odp} and constraints for the EoS of the dark energy~\cite{Aghanim:2018eyx}, we found that the anomaly factor needed for explaining the cosmic birefringence reads $c_\gamma \sim \mathcal{O}(20)$, which the model can successfully accommodate thanks to its UV structure. Importantly, combined with the theoretical requirement for the perturbativity of the model, the observational constraints produce $-0.994<w_{\rm DE,0}<-0.968$ (68\% C.L.) as the model's prediction for EoS of EW axion DE. This will improve as the near-future cosmic birefringence experiment becomes preciser. Upcoming probes will soon be able to test this scenario.

\vspace{-0.5cm}
\begin{acknowledgments}
The original idea of the present work arose during the TDLI International Workshop ``Current Topics on Axion'' on May 15th, 2021. T.T.Y.~is supported in part by the China Grant for Talent Scientific Start-Up Project and the JSPS Grant-in-Aid for Scientific Research No.~16H02176, No.~17H02878, and No.~19H05810 and by World Premier International Research Center Initiative (WPI Initiative), MEXT, Japan. L.V.~acknowledges support from the European Union's Horizon 2020 research and innovation programme under the Marie Sk{\l}odowska-Curie grant agreement No.~754496 (H2020-MSCA-COFUND-2016 FELLINI).
\end{acknowledgments}

\bibliographystyle{apsrev4-1}
\bibliography{ref}

\end{document}